\documentclass[aps,prl,twocolumn]{revtex4}
\usepackage{amssymb}
\usepackage{amsfonts}
\usepackage{amsmath}
\usepackage{graphicx}
\usepackage{epstopdf}
\usepackage{pstricks}
\usepackage{color}
\usepackage{hyperref}
\usepackage{pifont}
\usepackage{bm}

\usepackage{xcolor}
\definecolor{cinnamon}{rgb}{0.82, 0.41, 0.12}

\hypersetup{colorlinks=true,linkcolor=blue,citecolor=blue,urlcolor=blue} 

\newcommand{\ub}{{\bf u}}
\newcommand{\Ub}{{\bf U}}
\newcommand{\Xb}{{\bf X}}

\newcommand{\xb}{{\bf x}}
\newcommand{\fb}{{\bf f}}
\newcommand{\Fb}{{\bf F}}

\begin{document}

\title{A flexible fiber reveals the two-point statistical properties of turbulence}
\author{Marco Edoardo Rosti$^{1}$, Arash Alizad Banaei$^{1}$, Luca Brandt$^{1}$, Andrea Mazzino$^{2,3}$}
\affiliation{$^1$Linn\'{e} Flow Centre and SeRC (Swedish e-Science Research Centre), \\KTH Mechanics, SE 100 44 Stockholm, Sweden,\\
  $^2$DICCA, University of Genova, Via Montallegro 1, 16145 Genova, Italy\\
$^3$INFN and CINFAI Consortium,  Genova Section, Via Montallegro 1, 16145 Genova, Italy\\
}
\date{\today}

\begin{abstract}
We study the dynamics of a flexible fiber freely moving in a three-dimensional fully-developed turbulent field and present a phenomenological theory to describe the interaction between the fiber elasticity and the turbulent flow. This theory leads to the identification of two distinct regimes of flapping, which we validate against Direct Numerical Simulations (DNS) fully resolving the fiber  dynamics. The main result of our analysis is the identification of a flapping regime where the fiber, despite its elasticity, is slaved to the turbulent fluctuations. In this regime the fiber can be used to measure two-point statistical observables of turbulence, including scaling exponents of velocity structure functions, the sign of the energy cascade and the energy flux of turbulence, as well as the characteristic times of the eddies within the inertial range of scales. Our results are expected to have a deep impact on the experimental turbulence research as a new way, accurate and efficient, to measure two-point, and more generally multi-point, statistics of turbulence. 
\end{abstract}

\maketitle

Understanding how elastic structures interact with a turbulent flow is a problem attracting a great deal of attention in different fields of science and technology, ranging from biological applications  \cite{FL06,BMB12,Mc81,Bo12} to energy harvesting  \cite{Mc81,Bo12,Li16,Zh17}.

\begin{figure}
\centering
\vspace{-0.cm}
\includegraphics[keepaspectratio,width=7cm]{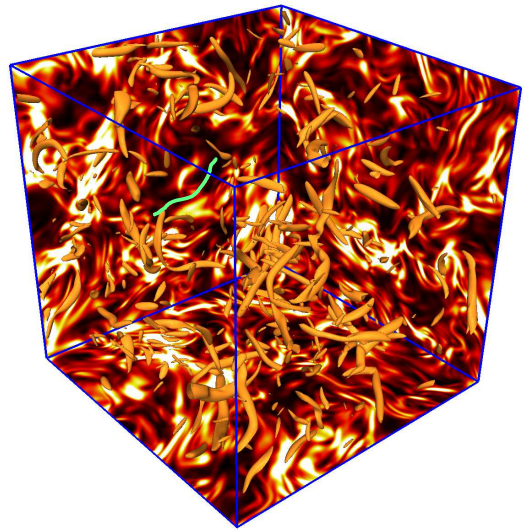} 
\caption{Visualization of a flexible fiber (green line) immersed in a homogeneous isotropic turbulent flow. The instantaneous vorticity field is represented by means of $\mathcal{Q}$-isosurfaces, while the three back planes show the contours of the enstrophy field.}
\label{fig:snap}
\end{figure}

The study in Ref.~\cite{ZCLS00} enabled a huge step forward in understanding the coupling between laminar flows and structure elasticity. This breakthrough was possible thanks to the combined choice of a simple flow configuration (a soap film used as a laminar two-dimensional flow tunnel \cite{Co89, Ma95, Ke95, NatCom14}) and a simple elastic structure (a flexible fiber of given rigidity and inertia).  Even in this apparently simple configuration the coupling between fluid and structure gives rise to a nontrivial and rich phenomenology. Once this has been described and the underlying mechanisms understood, new open questions arise about the dynamics of a fiber freely-moving in a three-dimensional turbulent environment (see Fig. \ref{fig:snap}): how does a flexible fiber interact with a turbulent flow? Under which conditions will flapping motion appear? How many states of flapping are possible? Can we control the amplitude/frequency of the resulting flapping states? Can the fiber be used to reveal the two-point statistics of turbulence?

Answering these questions is the main objective of the present Letter. Our findings will therefore also indicate how to exploit the motion of a flexible fiber in turbulence to obtain a proxy of two-point (and multi-point as a further generalization suggested at the end of the present Letter) statistics of turbulence.
A deep and complete understanding of turbulence, currently still missing, depends on the possibility of having accurate measurements of multipoint statistics (i.e.~measurements of simultaneous velocity correlation functions between different spatial points). These types of measurements are crucial for establishing a connection between scaling laws and spatial structures, e.g.~vortex filaments \cite{S81,DCB91}. Lagrangian particle tracking techniques helped us to successfully characterize Lagrangian statistics of turbulence \cite{XONB07} but a general mapping between Lagrangian and Eulerian statistics of turbulence still remains elusive. The main problem when using tracers to access Eulerian statistics of turbulence is that particles tend to separate from each other by virtue of the well-known Richardson law, which prevents obtaining converged statistics for a given fixed separation between the particles. Our aim here is to propose a new strategy where the concept of particle tracking is replaced by the new concept of fiber tracking. The idea is to exploit a fiber in a way never tried before, thus overcoming the problem related to single particle dispersion: the end-to-end distance of a fiber cannot indeed become larger than the fiber length at rest, making it  a good candidate to measure statistics on a given scale, i.e.~its end-to-end distance. A polidisperse, dilute, solution of fibers of different lengths should be considered to access all scales involved in a turbulent field.

Such a line of research is still in its infancy. A characterisation of the dynamics of an elastic fiber in turbulence has been recently provided in Refs.~\cite{Brouzet14, VB16} even if limited to the so-called over-damped regime (Le Gal, private communication). The idea to use small elastic objects (i.e.~deformable particles) to measure single-point velocity gradient was recently proposed in Ref.~\cite{APS}, even if no access to inertial-range multipoint measurement was considered. Our aim is to fill the gap with the final aim of proposing a new strategy to perform inertial-range measurements of turbulence two-point statistics by the Lagrangian tracking of an elastic fiber.

We tackle the problem at hand exploiting, in synergy, the phenomenological theory we propose here, and numerical simulations fully resolving the fiber dynamics  in three-dimensional homogeneous isotropic stationary turbulence (see Fig.~\ref{fig:snap}).

First, we present the model coupling the fiber dynamics and the flow. The fluid flow, $\ub(\xb,t)$, is governed by the mass and momentum conservation equations, written including a fluid-structure interaction force $\fb$ \cite{PSK02,PSK07,HSS07},
\begin{equation}
\begin{array}{lcl}\vspace{0.2cm}
  \partial_t \ub + \ub\cdot\bm{\partial}\ub  		&=& 	-\bm{\partial} p/\rho_0 +
  \nu\partial^2\ub +  \fb,\\\vspace{0.2cm}
\bm{\partial}\cdot\ub &=& 0,
\label{eq:1}
\end{array}
\end{equation}
while the fiber position, $\Xb(s,t)$, is governed by the Euler-Bernoulli beam equation and by the inextensibility constraint \cite{S07}
\begin{equation}
\begin{array}{lcl}\vspace{0.2cm}
\rho_1\ddot\Xb &=& \partial_s(T \partial_s (\Xb)) - \gamma\partial^4_{s}(\Xb) + \Fb, \\\vspace{0.2cm}
\partial_{s}(\Xb) \cdot \partial_{s}(\Xb) &=& 0.
\label{eq:2}
\end{array}
\end{equation}
In the previous set of equations, $s$ is the curvilinear abscissa, $\rho_0$ and $\nu$ are the fluid density and kinematic viscosity, $\rho_1$ the difference between the linear density of the fiber and fluid, $\gamma$ the fiber bending rigidity (for a homogeneous fiber, it is the product of the elastic modulus and the second moment of area), and $T$ is the tension needed to enforce the fiber inextensibility. The fluid and the fiber are coupled at their interface by the no-slip condition $\dot\Xb=\Ub(\Xb(s,t),t)$, with $\Ub(\Xb(s,t),t) = \int \ub(\xb,t)\delta(\xb-\Xb(s,t))\ d\xb$ the Lagrangian fiber velocity and $\fb(\xb,t) = \int_s\Fb(s,t)\delta(\xb-\Xb(s,t))\ ds$, where $\fb(\xb,t)$ is the Eulerian fluid-structure interaction force density and $\Fb(s,t)$ the Lagrangian force density. Free-end conditions are used at $s=0$ and $s=c$, $c$ being the rest length of the fiber. An additional volume force is considered on the right hand side of the Navier--Stokes equations in (\ref{eq:1}) (not shown for the sake of brevity) to generate a fully-developed turbulent state with isotropic, homogeneous, and stationary statistics. 

Before investigating numerically the fully-coupled problem, let us start the analysis by focusing on the fiber equation in a given turbulent environment obeying the well-known Kolmogorov theory \cite{F95}. Such an intrinsically  passive way of thinking at the fiber dynamics has an analog in polymer physics \cite{BC18,PRB08} where it was successful, e.g., to predict the statistics of polymer elongations in a turbulent flow \cite{BMM11}. To describe the fluid-structure interaction, let us assume a viscous coupling of the form $ \Fb=-\mu (\dot \Xb - \ub)$, with $\mu$ being the dynamic viscosity of the flow \cite{Cox70}. Note that here we do not consider an anisotropic expression for the drag, as done e.g. in Ref.~\cite{YS07} for passive fibers in small Reynolds number flows.  We choose to not complicate the description given the intrinsically isotropic nature of the underlying turbulence flow which causes no preferential alignment. Indeed, this simple isotropic description is able to properly describe the fiber dynamics, as shown below. On this basis, two characteristics time-scales can be immediately identified from the fiber equation: the viscous time-scale $\tau_{\mu}={2\rho_1}/{\mu}$ (obtained by balancing fiber inertia with  viscous damping) and the fiber elastic time $\tau_{\gamma}=\alpha ( {\rho_1c^4}/{\gamma})^{1/2}$ (obtained by balancing fiber inertia with bending rigidity) \cite{note3}. Different regimes are expected depending on the value of the {\it damping ratio} $\zeta={\tau_{\mu}}/{\tau_{\gamma}}=(\alpha c^2\mu)/(2\rho_1^{1/2}\gamma^{1/2})$. For $0<\zeta<1$ (under-damped regime) the elasticity is expected to strongly affect the fiber dynamics, while for $\zeta>1$ (over-damped regime) elastic effects  are strongly inhibited.

Let us focus on the former, dynamically richer, regime and start to analyze two opposite situations. For large elasticity, only large strains may appreciably deform the fiber, and when these events occur the fiber rapidly reacts trying to restore the straight position; the relaxation process is expected to be dominated by rapid oscillations of the characteristic elastic time $\tau_{\gamma}$. In the opposite limit, small fiber elasticity, the fiber does not resist deformation and is thus slaved to the turbulent fluctuations. We thus argue the existence of a critical value $\gamma_{\mbox{\tiny{crit}}}$ of the fiber bending rigidity separating these two distinct behaviors and claim that $\gamma_{\mbox{\tiny{crit}}}$ can be extracted from a resonance condition between the fiber elastic time $\tau_{\gamma}$ and the eddy turnover time $\tau(r)=r^{2/3}{\epsilon}^{-1/3}$ evaluated at the fiber scale $c$, $\epsilon$ being  the turbulence dissipation rate of kinetic energy. The condition $\tau(c)\sim \tau_{\gamma}$ immediately gives: 
\begin{equation}
\alpha\left ( \frac{c^4 \rho_1}{\gamma}\right )^{1/2} \sim c^{2/3}\epsilon^{-1/3}\to \gamma_{\mbox{\tiny{crit}}}\sim c^{8/3}\epsilon^{2/3} \rho_1 \alpha^2 \;\;\;.
\label{gamma-crit}
\end{equation}

\begin{figure}
\centering
\vspace{-0.cm}
\includegraphics[keepaspectratio,width=8cm]{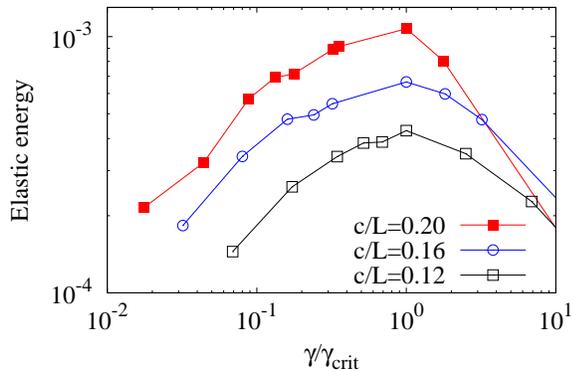} 
\caption{The fiber elastic energy made dimensionless with $1/2\rho_0\int u^2d\xb$ for three different fiber lengths. $L$ denotes the size of the computational domain. The peaks at $\gamma\sim \gamma_{\mbox{\tiny{crit}}}$ are the fingerprint of a resonance between the fiber elastic time scale and the eddy turnover time evaluated at the fiber length scale.}
\label{fig:reson}
\end{figure}

The remaining of the present Letter is devoted to prove our conjectures exploiting accurate DNS coupled to an efficient IBM strategy to resolve the fully-coupled fiber-flow dynamics. Details on the numerical strategy are given in Refs.~\cite{RB17,note1,HSS07}.

To start, let us provide a justification for the term `resonance' we have associated to the condition (\ref{gamma-crit}). In Fig.~\ref{fig:reson} we report the fiber elastic energy as a function of $\gamma/\gamma_{\mbox{\tiny{crit}}}$ for different values of the fiber length in the under-damped regime ($0<\zeta<1$). The peaks at $\gamma/\gamma_{\mbox{\tiny{crit}}}\sim 1$ provide a first clue that $\gamma_{\mbox{\tiny{crit}}}$ plays a dynamical role.

\begin{figure}
\centering
\vspace{-0.cm}
\includegraphics[keepaspectratio,width=8cm]{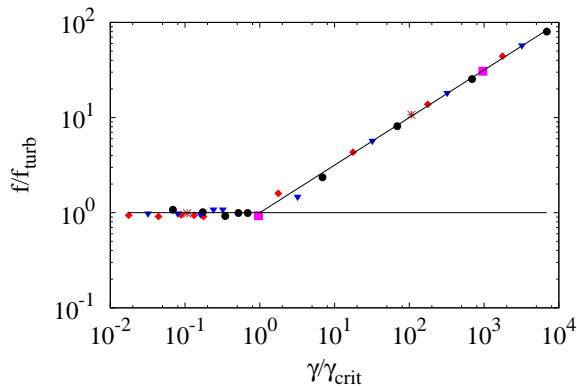} 
\caption{The fiber oscillation frequencies (normalized by the turbulence frequency at the fiber length scale) as a function of the fiber bending rigidity (normalized by the critical value given in (\ref{gamma-crit})). Diamonds: $c/L=0.20$ and $\rho_1/(\rho_0 L^2)=0.042$; bullets: $c/L=0.12$ and $\rho_1/(\rho_0 L^2)=0.042$; triangles: $c/L=0.16$ and $\rho_1/(\rho_0 L^2)=0.042$; stars: $c/L=0.16$ and $\rho_1/(\rho_0 L^2)=0.125$; squares: $c/L=0.16$ and $\rho_1/(\rho_0 L^2)=0.014$.}
\label{fig:slaved}
\end{figure}

To identify the role of $\gamma_{\mbox{\tiny{crit}}}$ we have analyzed long time-series (corresponding to $\sim 20$ large-eddy turnover times), of the motion of $30$ different fibers, corresponding to different combinations of $3$ different densities $\rho_1$, $3$ lengths $c$ and $9$ bending rigidities $\gamma$, all the cases belonging to the under-damped regime. The leading oscillation frequency $f$, extracted from the Fourier transform of the time history of the end-to-end distance and divided by $f_{turb}=1/\tau(c)$, is reported in Fig.~\ref{fig:slaved} as a function of $\gamma/\gamma_{\mbox{\tiny{crit}}}$. The outcome confirms our expectations and the good data collapse brings to the following three main conclusions: i) $\gamma_{\mbox{\tiny{crit}}}$ separates two distinct regimes in the under-damped case with a sharp transition; ii) for $\gamma<\gamma_{\mbox{\tiny{crit}}}$, the most energetic mode of oscillation of the fiber is at the turbulence frequency $1/\tau(c)$; iii) for $\gamma>\gamma_{\mbox{\tiny{crit}}}$, the most energetic mode of oscillation is associated to the first fiber normal mode and has the frequency $1/\tau_{\gamma}$. The interested reader is referred to the Supplemental Material \cite{supp} for more information on the fiber dynamics.

The fact that for $\gamma/\gamma_{\mbox{\tiny{crit}}}<1$ the fiber is locked to the frequency of the turbulent eddies with the same size of the fiber suggests 
that the fiber is able to reveal the turbulence velocity fluctuations. In plain words, we consider our fiber as a physical proxy of the celebrated turbulent eddies. This being the case, a massive fiber, which can be easily tracked in a turbulent flow, may reveal the features of eddies of different scales. To demonstrate that our conjecture is true, we compute the longitudinal structure functions $S_p(r)$, $p=2,3$, defined in terms of the fiber velocities at the fiber end points projected along the end-to-end fiber vector for $\gamma=\gamma_{\mbox{\tiny{crit}}}/2$. We  compute $S_p(r)$ from three different fibers, with different rest lengths belonging to the inertial range of scales. As far as the separation $r$ is concerned, instead of the fiber rest length, we use the time-averaged value of the fiber end-to-end distance, as this is a quantity more representative of the dynamical fiber length. All regimes analyzed fall in the under-damped case and each fiber has been tracked for $40$ large-eddy turnover times.
\begin{figure}
\centering
\vspace{-0.cm}
\includegraphics[keepaspectratio,width=8cm]{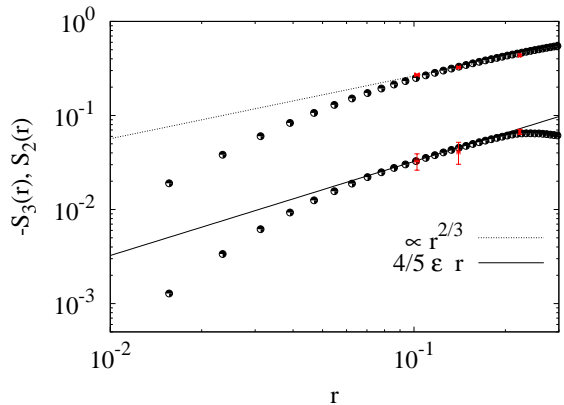} 
\caption{Second-order and third-order velocity structure functions for $\gamma/\gamma_{\mbox{\tiny{crit}}}=1/2$. Black bullets represent the standard Eulerian measure while symbols with error bars are the results obtained from the fibers as detailed in the main text. Lengths and velocities are made dimensionless with the box size $L$ and with the velocity root mean square, respectively.}
\label{fig:sfun}
\end{figure}
The results are presented in Fig.~\ref{fig:sfun}, where the second and third-order velocity structure functions obtained by the fiber motion are compared to those 
obtained following the standard Eulerian procedure, with averages both in space and in time 
given the homogeneity and stationarity of the turbulence statistics (black bullets). These convincingly show 
the celebrated Kolmogorov 4/5th law for the third-order structure function. The markers indicate the structure functions obtained from the fibers, and the errorbars have been determined from the convergence profile of both structure functions (ordinates) and end-to-end fiber distances (abscissa). The agreement between the Eulerian measurements and those obtained from the fibers is within error bars. Note that, the value of $\epsilon$ used here has been determined independently from its definition, and there are thus no free parameters. 
 
\begin{figure}
\centering
\vspace{-0.cm}
\includegraphics[keepaspectratio,width=8cm]{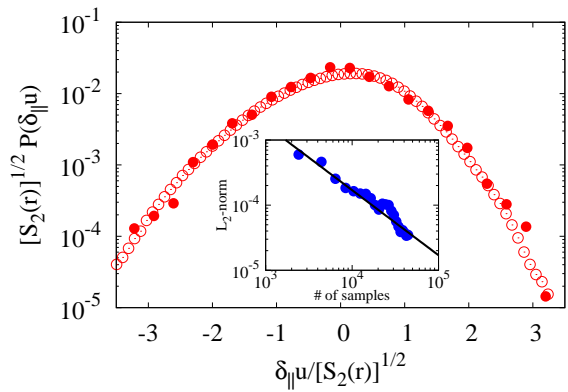} 
\caption{The probability density function (pdf) of velocity increments for a separation between points belonging to the inertial range of scales ($r= 0.14 L$, corresponding to the second marker with error bars of Fig.~\ref{fig:sfun}). Open circles: Eulerian pdf; bullets: pdf from the fiber.
 Inset: behavior of the $L^2$-norm of the difference between the Eulerian and the fiber-based pdf. The continuous line corresponds to the $N^{-1}$ behavior, $N$ being the number of fiber-based samples.}
\label{fig:pdf}
\end{figure}
A further confirmation comes from Fig.~\ref{fig:pdf} where we have reported the probability density function (pdf) of longitudinal velocity increments for a separation corresponding to one of the three fibers reported in Fig.~\ref{fig:sfun}. Open circles depict the pdf obtained in the Eulerian frame (about $6\times 10^6$ samples considered) while bullets indicate the pdf from the longitudinal velocity differences evaluated from the fiber velocities at the fiber end points (about $5\times 10^4$ samples). The small discrepancy among the two pdf can be associated to the lack of statistics in the Lagrangian description (which is about a factor 100 smaller than that of the Eulerian frame). However, the agreement  increases by increasing the number of statistical samples of the fiber-based measurements as shown in the inset of Fig.~\ref{fig:pdf} where the $L^2$ norm of the difference between the two pdf is shown as a function of the number $N$ of statistical samples. The error approximately decreases as $1/N$. Similar agreement has been observed (not shown) for the other two fiber lengths considered in Fig.~\ref{fig:sfun}. We can thus conclude that choosing  $\gamma<\gamma_{\mbox{\tiny{crit}}}$ allows one to measure the eddy turnover time of turbulence at the fiber length-scale, and to quantitatively access the statistical properties of the two-point statistics of turbulence.

There remain to discuss the over-damped regime ($\zeta>1$), when the fiber equation becomes first-order in time: once deformed, the fiber reacts exponentially with the typical time scale $\mu c^4 /\gamma$ and no elastic oscillations occur. For $\mu c^4 /\gamma \ll \tau (c)$ the relaxation process is faster than the eddy turnover time at the length-scale of the fiber, while the opposite occurs for $\mu c^4 /\gamma \gg \tau (c)$. A critical value of the fiber bending rigidity separating the two regimes can thus be identified: $\gamma^{od}_{\mbox{\tiny{crit}}}\sim \mu c^{10/3}\epsilon^{1/3} $. For different reasons, we argue that the fiber undergoes oscillations with frequency $\sim 1/\tau(c)$ in both limits. For $\gamma/\gamma^{od}_{\mbox{\tiny{crit}}}<1$ all points of the fiber are indeed expected to move with the fluid velocity under the constraint of fiber inextensibility. For $\gamma/\gamma^{od}_{\mbox{\tiny{crit}}}>1$, as in the under-damped case, only large strains may deform the fiber, and the fiber rapidly reacts back trying to restore the straight position. The relaxation process takes place without oscillations and we thus expect that, differently from the under-damped regime, oscillations have frequency $\sim 1/\tau(c)$. Our expectation has been verified numerically for $\rho_1  \sim 0$, corresponding to $\gamma/\gamma^{od}_{\mbox{\tiny{crit}}}\gg 1$ and the results (not shown) fully confirm our guess. Note that, Refs.~\cite{Brouzet14, VB16} provide a slightly different expression for $\gamma^{od}_{\mbox{\tiny{crit}}}$, i.e.~$\gamma^{od}_{\mbox{\tiny{crit}}}\sim c^4 (\rho_0 \mu\epsilon)^{1/2}$. A possible reason for the discrepancy between the two formulations is that the fibers considered in that references are close to the integral scale of the flow while they are well within the inertial range of scales in the present case.

In conclusion, we have explored the dynamical properties of a single elastic fiber with length within the inertial range of scales, free to evolve in a  turbulent field. The main result of our analysis has been the identification of a dynamical regime where the fiber, in spite of its elasticity, is slaved to turbulence thus becoming a material realization of the well-known concept of turbulent eddy. Our results extends to inertial range two-point statistics the idea of using deformable particles for single particle measurements of velocity gradient recently presented in Ref.~\cite{APS}. Further pioneering extensions to multi-point statistics in turbulence seem to be realizable exploiting flexible membranes or other spatially-extended elastic objects.

\section*{Acknowledgment}
\noindent A.M. thanks useful discussions with A. Cauteruccio, S. Olivieri and S. Putzu during the 2018 Ph.D Modelling Camp held at the Genova University. Useful discussions during the COST meeting Flowing Matter 2018 (Lisbon, Portugal) are also acknowledged. M.E.R, A.A.B and L.B. were supported by the ERC-2013-CoG-616186 TRITOS, and by the VR 2014-5001. The authors acknowledge the computer time provided
by SNIC and INFN-CINECA.


\begin{thebibliography}{99}
\bibitem{FL06} F. Fish and G. Lauder,
Ann. Rev. Fluid Mech. 38, 193–224 (2006)

\bibitem{BMB12} S. Bagheri, A. Mazzino and A. Bottaro,
Phys. Rev. Lett. 109, 154502 (2012)

\bibitem{Mc81} W. McKinney and J. DeLaurier,
J. Energy, 109 (1981) 

\bibitem{Bo12} C. Boragno, R. Festa and A. Mazzino,
Appl. Phys. Lett., 253906 (2012)

\bibitem{Li16} D. Li, Y. Wu, A. Da Ronch and J. Xiang,
Prog. Aerosp. Sci. 86, 28 (2016)

\bibitem{Zh17} L. Zhao and Y. Yang,
Shock and Vibration 2017, 31  (2017)

\bibitem{ZCLS00} J. Zhang, S. Childress, A. Libchaber and M. Shelley,
Nature {\bf 408}, 835 (2000)

\bibitem{Co89} Y. Couder, J.M. Chomaz and M. Rabaud,
Physica D 37, 384 (1989)

\bibitem{Ma95} B.K. Martin and X.-L. Wu,
Rev. Sci. Instrum. 66, 5603 (1995)

\bibitem{Ke95} H. Kellay, X.-L. Wu and W.I.  Goldburg,
Phys. Rev. Lett. 74, 3975 (1995)

\bibitem{NatCom14} U. L\={a}cis, N. Brosse, F. Ingremeau, A. Mazzino, F. Lundell, H. Kellay and S. Bagheri,
Nature Communications 5, 5310 (2014) 

\bibitem{S81} E. D. Siggia,
J. Fluid Mech. 107, 375 (1981)

\bibitem{DCB91} S. Douady, Y. Couder and M. E. Brachet,
Phys. Rev. Lett. 67, 983 (1991)

\bibitem{XONB07} H. Xu, N. T. Ouellette, H. Nobach and E. Bodenschatz,
Advances in Turbulence XI (2007)

\bibitem{Brouzet14} C. Brouzet, G. Verhille and P. Le Gal,
Phys. Rev. Lett 112, 074501  (2014)

\bibitem{VB16} G. Verhille and A. Bartoli,
Exp. Fluids 57:117 (2016)

\bibitem{APS} B. Hejazi, M. Krellenstein and G. Voth,
Bull. Am. Phy. Soc. 62 (2017)
  

\bibitem{HSS07} W. -X. Huang, S. J. Shin and H. J. Sung,
J. Comp. Phys. 226 2206 (2007).

\bibitem{PSK02} C.S. Peskin,
Acta Numerica {\bf 11}, 479 (2002)

\bibitem{PSK07} Y. Kim and C.S. Peskin,
Phys. Fluids {\bf 19}, 053103 (2007)

\bibitem{S07} L. A. Segel,
 SIAM 52 (2007)

\bibitem{F95} U. Frisch, Turbulence: The Legacy of AN Kolmogorov (Cambridge
University Press, Cambridge, England, 1995).
 
\bibitem{PRB08} I. Procaccia, V.S. L'vov and R. Benzi,
Rev. Mod. Phys. 80, 225 (2008)

\bibitem{BC18}  R. Benzi and E. Ching,
Ann. Rev. Cond. Mat. Phys. 9 (2018) 

\bibitem{BMM11}  G. Boffetta, A. Mazzino and S. Musacchio,
Physical Review E 83, 056318 (2011)

\bibitem{Cox70} R.G. Cox,
J. Fluid Mech. 44, 791  (1970)

\bibitem{YS07} Y. N. Young and M. J. Shelley
Phys. Rev. Lett. 99, 058303 (2007)

\bibitem{note3} The factor $\alpha=\pi/22.3733 $ follows from a standard normal mode analysis carried out on the fiber equation with free-end conditions and is associated to the natural oscillation frequency corresponding to the first mode of oscillation.  
 
 \bibitem{note1} The Navier--Stokes equations are solved using a second order finite-difference scheme in space and third-order Runge-Kutta scheme in time. The pressure is obtained by solving the Poisson equation using Fourier transforms. We use a Cartesian uniform mesh in a rectangular triperiodic box of size $L=2\pi$, with 128 grid points per side. The grid size is sufficient to obtain a clear inertial range of scale clearly displaying the expected $4/5th$ Kolmogorov law. Doubling the resolution in all directions results in an insignificant change in the results. The turbulent dissipation rate $\epsilon$, made dimensionless with the cube of the velocity root-mean square divided by the size of the box, is $2.54$ and the Reynolds number at the Taylor microscale is $Re_\lambda=92$. In order to sustain the turbulent field, we use the spectral forcing scheme described in Ref.~\cite{EP88}.

\bibitem{EP88} V. Eswaran and S. B. Pope,
Comput. Fluids 16, 3, (1988)

\bibitem{RB17} M. E. Rosti and L. Brandt,
J. Fluid Mech. 830, 708 (2017) 

\bibitem{supp} See Supplemental Material at [URL will be inserted by publisher] to watch movies showing the fiber dynamics in the under-damped regime for $\gamma<\gamma_{\mbox{\tiny{crit}}}$ and for $\gamma>\gamma_{\mbox{\tiny{crit}}}$.

\end{thebibliography}
\end{document}